\documentclass[twoside,twocolumn,9pt]{article}
\usepackage{extsizes}
\usepackage[super,sort&compress,comma]{natbib} 
\usepackage[version=3]{mhchem}
\usepackage[left=1.5cm, right=1.5cm, top=1.785cm, bottom=2.0cm]{geometry}
\usepackage{balance}
\usepackage{mathptmx}
\usepackage{sectsty}
\usepackage{graphicx} 
\usepackage{lastpage}
\usepackage[format=plain,justification=justified,singlelinecheck=false,font={stretch=1.125,small,sf},labelfont=bf,labelsep=space]{caption}
\usepackage{float}
\usepackage{fancyhdr}
\usepackage{fnpos}
\usepackage[english]{babel}
\addto{\captionsenglish}{%
  
}
\usepackage{array}
\usepackage{droidsans}
\usepackage{charter}
\usepackage[T1]{fontenc}
\usepackage[usenames,dvipsnames]{xcolor}
\usepackage{setspace}
\usepackage[compact]{titlesec}
\usepackage{hyperref}

\usepackage{epstopdf}

\definecolor{cream}{RGB}{222,217,201}

\begin{document}

\pagestyle{fancy}
\thispagestyle{plain}
\fancypagestyle{plain}{
\renewcommand{\headrulewidth}{0pt}
}

\makeFNbottom
\makeatletter
\renewcommand\LARGE{\@setfontsize\LARGE{15pt}{17}}
\renewcommand\Large{\@setfontsize\Large{12pt}{14}}
\renewcommand\large{\@setfontsize\large{10pt}{12}}
\renewcommand\footnotesize{\@setfontsize\footnotesize{7pt}{10}}
\makeatother

\renewcommand{\thefootnote}{\fnsymbol{footnote}}
\renewcommand\footnoterule{\vspace*{1pt}%
\color{cream}\hrule width 3.5in height 0.4pt \color{black}\vspace*{5pt}} 
\setcounter{secnumdepth}{5}

\makeatletter 
\renewcommand\@biblabel[1]{#1}            
\renewcommand\@makefntext[1]%
{\noindent\makebox[0pt][r]{\@thefnmark\,}#1}
\makeatother 
\renewcommand{\figurename}{\small{Fig.}~}
\sectionfont{\sffamily\Large}
\subsectionfont{\normalsize}
\subsubsectionfont{\bf}
\setstretch{1.125} 
\setlength{\skip\footins}{0.8cm}
\setlength{\footnotesep}{0.25cm}
\setlength{\jot}{10pt}
\titlespacing*{\section}{0pt}{4pt}{4pt}
\titlespacing*{\subsection}{0pt}{15pt}{1pt}

\fancyfoot{}
\fancyfoot[RO]{\footnotesize{\sffamily{1--\pageref{LastPage} ~\textbar  \hspace{2pt}\thepage}}}
\fancyfoot[LE]{\footnotesize{\sffamily{\thepage~\textbar\hspace{3.45cm} 1--\pageref{LastPage}}}}
\fancyhead{}
\renewcommand{\headrulewidth}{0pt} 
\renewcommand{\footrulewidth}{0pt}
\setlength{\arrayrulewidth}{1pt}
\setlength{\columnsep}{6.5mm}
\setlength\bibsep{1pt}

\makeatletter 
\newlength{\figrulesep} 
\setlength{\figrulesep}{0.5\textfloatsep} 

\newcommand{\topfigrule}{\vspace*{-1pt}%
\noindent{\color{cream}\rule[-\figrulesep]{\columnwidth}{1.5pt}} }

\newcommand{\botfigrule}{\vspace*{-2pt}%
\noindent{\color{cream}\rule[\figrulesep]{\columnwidth}{1.5pt}} }

\newcommand{\dblfigrule}{\vspace*{-1pt}%
\noindent{\color{cream}\rule[-\figrulesep]{\textwidth}{1.5pt}} }

\makeatother

\twocolumn[
  \begin{@twocolumnfalse}
\vspace{3cm}
\sffamily
\begin{tabular}{m{4.5cm} p{13.5cm} }

& \noindent\LARGE{\textbf{Interface-induced hysteretic volume phase transition of microgels: simulation and experiment}
} \\
\vspace{0.3cm} & \vspace{0.3cm} \\

 & \noindent\large{Jannis Kolker,$^{\ast}$\textit{$^{a}$} Johannes Harrer,\textit{$^{\ddag b}$} Simone Ciarella,\textit{$^{c}$} Marcel Rey,\textit{$^{b}$} Maret Ickler,\textit{$^{b}$} Liesbeth M. C. Janssen,\textit{$^{c}$} Nicolas Vogel,\textit{$^{b}$} and Hartmut L\"owen\textit{$^{a}$}} \\
\\
 & \noindent\normalsize{Thermo-responsive microgel particles can exhibit a drastic volume shrinkage upon increasing the solvent temperature.  Recently we found that the spreading of poly(N-isopropylacrylamide)(PNiPAm) microgels at a liquid interface under the influence of surface tension hinders the  temperature-induced volume phase transition. In addition, we observed a hysteresis behavior upon temperature cycling, i.e.\ a different evolution in microgel size and shape depending on whether the microgel was initially adsorbed to the interface in expanded or collapsed state. Here, we model the volume phase transition of such microgels at an air/water interface by  monomer-resolved Brownian dynamics simulations and compare the observed behavior with experiments. We reproduce the experimentally observed hysteresis in the microgel dimensions upon temperature variation. Our simulations did not observe any hysteresis for microgels dispersed in the bulk liquid, suggesting that it results from the distinct interfacial morphology of the microgel adsorbed at the liquid interface. An initially collapsed microgel brought to the interface and subjected to subsequent swelling and collapsing (resp.\ cooling and heating) will end up in a larger size than it had in the original collapsed state. Further temperature cycling, however, only shows a much reduced  hysteresis, in agreement with our experimental observations. We attribute the hysteretic behavior to a kinetically trapped initial collapsed configuration, which relaxes upon expanding in the swollen state. We find a similar behavior for linear PNiPAm chains adsorbed to an interface. Our combined experimental - simulation investigation provides new insights into the volume phase transition of PNiPAm materials adsorbed to liquid interfaces.}

\end{tabular}

 \end{@twocolumnfalse} \vspace{0.6cm}

  ]

\renewcommand*\rmdefault{bch}\normalfont\upshape
\rmfamily
\section*{}
\vspace{-1cm}


\footnotetext{\textit{$^{a}$~
Institut fur Theoretische Physik II, Heinrich-Heine-Universit\"at
D\"usseldorf, Universit\"atsstrasse 1, D-40225 D\"usseldorf, Germany.}}
\footnotetext{\textit{$^{b}$~ Institute of Particle Technology, Friedrich-Alexander University Erlangen-N\"urnberg, Cauerstrasse 4, 91058 Erlangen, Germany. }}
\footnotetext{\textit{$^{c}$~Theory of Polymers and Soft Matter, Department of Applied Physics, Eindhoven University of Technology, P.O. Box 513, 5600MB Eindhoven, The Netherlands. }}


\pagebreak
\section{Introduction}
Soft microgels consisting of a swollen polymer network possess fascinating properties. Their soft nature allows them to undergo significant deformations, for example when adsorbed to an interface~\cite{Geisel2012,Rey2020}. In addition, stimuli-responsive properties can be encoded in their molecular structure, allowing microgels to undergo sharp transitions between swollen and collapsed states upon an external stimulus~\cite{Pelton2000,Karg2019,Keidel2018}.  This dynamic behavior has important consequences. From an applied side, stimuli-responsive microgels serve as dynamic and reversible emulsion stabilizers~\cite{Li2013,Ngai2005,Schmitt2013,fernandez2020microgels,TATRY202196} or vehicles for drug-delivery, due to their ability to take up and release molecules on demand~\cite{klinger2012stimuli,Karg2019}. From a fundamental side, soft microgels serve as model systems for classical many-body systems to model atomistic physical phenomena such as melting of crystal lattices~\cite{Wang2012,alsayed2005premelting}, solid-solid phase transitions~\cite{peng2015}, or complex self-assembly~\cite{Grillo2020, Ciarella2020,volk2019moire}.

Among the most prominent microgel systems are crosslinked poly(N-isopropylacrylamide) (PNiPAm)-based microgels, which exhibit a drastic volume shrinkage in aqueous dispersion above $32^\circ\text{C}$ ~\cite{Lee1999,Senff1999,Shiraga2015}.
During this volume phase transition, nonpolar groups of the PNiPAm microgel aggregate and bound water is expelled
from the macromolecule to increase the entropy of the
system~\cite{Pelton2000,Senff1999,Shiraga2015}. This volume phase transition has been extensively explored through experimental studies~ \cite{Shiraga2015,Keidel2018,Conley2016,Kratz2001}, computer simulations ~\cite{rovigatti2017internal,camerin2018modelling,Gnan2017,DelMonte2019,moreno2018computational,Keidel2018,weyer2018concentration,quesada2012computer,brugnoni2018swelling,scotti2019deswelling,nikolov2018mesoscale,denton2016counterion,hedrick2015structure,giupponi2007polyelectrolyte} and statistical theory ~\cite{hertle2010responsive,Lopez2017,Zhi2010,Toh2014,baul2020structure, Lin2020,MonchoJorda2016,giupponi2004monte,}. 

However, in many actual applications such as emulsion stabilization~\cite{Li2013,Ngai2005,Schmitt2013} or surface patterning~\cite{Goerlitzer2020,Rey2016,Grillo2020} microgel particles are exposed and attracted to a liquid interface, where the spatial isotropy of the microgel is broken and a more complex behavior results. Therefore it is important to understand their structural and dynamical  properties when confined to a liquid interface. In fact, it was shown that at liquid interfaces, microgels deform into a fried-egg-like structure ~\cite{Geisel2012,Style2015,Mehrabian2016,Zielinska2016,Rey2020} where the core possesses a disk-like shape and the surrounding dangling polymer chains form a quasi two-dimensional layer close to the interface, termed the corona. Correspondingly, the volume phase transition is strongly modified due to the presence of the interface. Recent experiments ~\cite{Harrer2019,Bochenek2019,bochenek2020temperature} have analyzed the volume phase transition at the interface and showed that the volume phase transition in the lateral direction is strongly hindered by the presence of the liquid interface. In addition, a hysteresis behaviour upon swelling and deswelling of microgels dependent on whether the microgels were deposited in the swollen or collapsed state was found~\cite{Harrer2019,Li2014}. 
Though  monomer-resolved computer simulations have considered the microgel structure at an interface~ \cite{camerin2020microgels,camerin2019microgels,Harrer2019}, the  hysteretic behaviour was not yet reproduced in computer simulations. 

\begin{figure}[h]
\includegraphics[height=3.5cm]{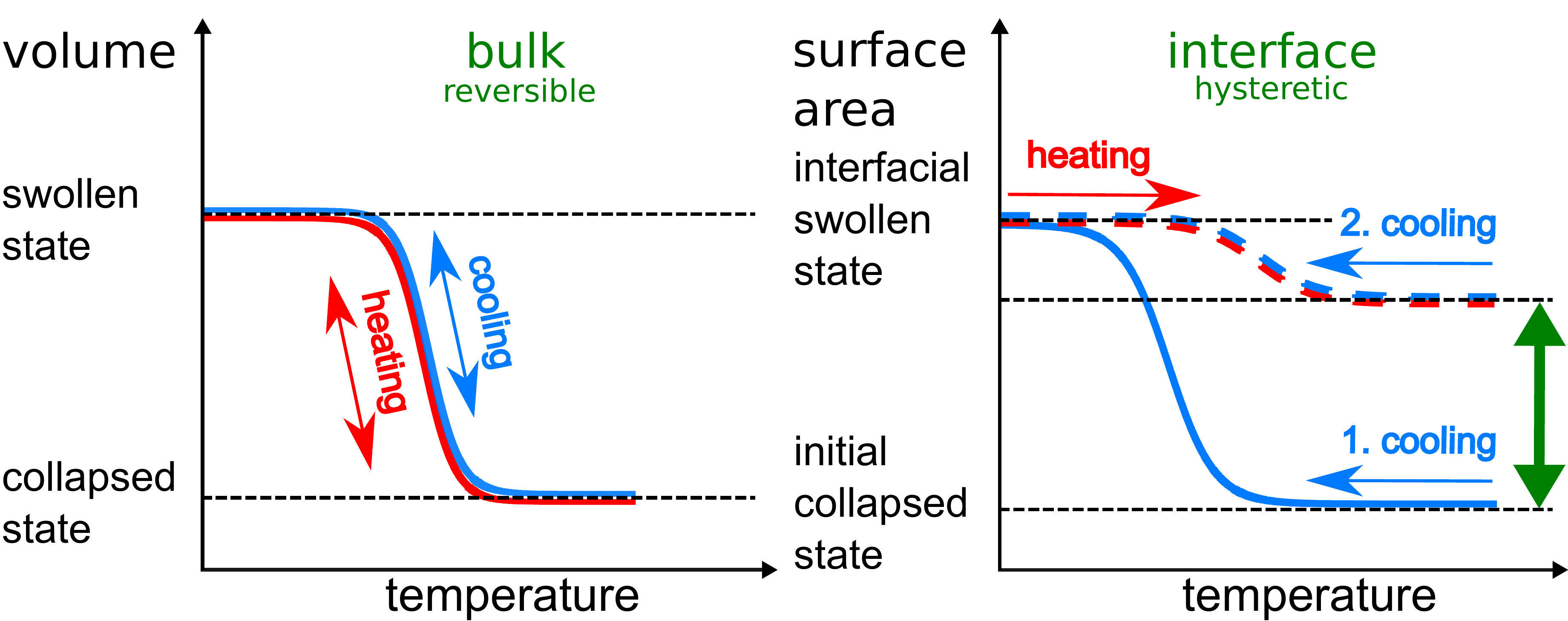}
\caption{Schematic visualization of the interface-induced hysteresis effect (right) compared to the  bulk behavior (left). Bulk:  reversibility between the swollen and collapsed state within a cooling-heating cycle for a microgel particle. Interface: hysteresis in the first cooling-heating cycle of an interface-adsorbed microgel particle. For subsequent cycles reversibility occurs. }
\label{fig:cartoon}
\end{figure}

Here we present extensive Brownian dynamics simulations to confirm and investigate the hysteresis on a monomer-resolved scale. The microgel particle is modelled by a coarse-grained polymer network with effective beads~\cite{rovigatti2017internal,camerin2018modelling}. The interface is described by an external effective potential attractive for the beads. Within our simulations we are able to reproduce and quantify the hysteresis behavior and the corresponding structural change of the microgel particle. In our protocol the microgel particle is first equilibrated in the bulk either in the collapsed (high temperature) or in the  swollen (low temperature) state and then brought to the interface and "equilibrated" there again. Figure ~\ref{fig:cartoon} schematically outlines the temperature cycling experiments used for this investigation. A swollen particle adsorbed to the interface responds reversibly to further collapsing and swelling (resp.\ heating and cooling) cycles. An initially collapsed particle adsorbed to the interface, however, experiences structural changes upon temperature cycling. Instead of reversible regaining its initial size, the microgel remains significantly more expanded after undergoing a temperature cycle. We attribute this hysteretic behaviour to an initial kinetically trapped collapsed state. This behavior is purely induced by the interface, as a similar temperature cycling of a bulk microgel does not show any hysteresis. We investigate these interfacial volume phase transitions as a function of crosslinker density. In both experiments and simulations, we find that less crosslinked microgels exhibit a larger hysteresis. In addition, we show that even linear PNiPAm chains adsorbed to an interface undergo a hysteretic behavior upon temperature cycling, indicating that the hysteresis does not only depend on the architecture of the microgel, but originates from the molecular nature of the polymer itself.  

\section{\label{II}Materials and Methods}
We adopt the modelling proposed in Ref.\ ~\cite{Gnan2017,camerin2019microgels,bergman2018new,rovigatti2019connecting} describing a microgel particle  on a monomer-resolved level with no explicit solvent. 
 In detail, the microgel particle consists of  two types of beads, monomers and crosslinkers, which define its internal architecture. A monomer is covalently linked to either a neighbouring other monomer or to a crosslinker by springs and the maximum number of these bonds is two. A crosslinking bead, on the other hand, has  four of such bonds. In terms of all other interactions (next nearest neighbour interactions etc.) monomers and crosslinkers do not differ. 
The covalent bonds are described by a finite-extensible-nonlinear-elastic (FENE) potential~\cite{Gnan2017,camerin2019microgels} (see
Supporting Information) with a characteristic energy scale $\varepsilon$, a maximal bond expansion $\tilde{R}_0=1.5\sigma$ and an effective spring constant $\tilde{k}_f=15\varepsilon/\sigma^2 $. The remaining bead-bead interactions are modelled by a repulsive Week-Chandler-Andersen (WCA) potential\cite{Gnan2017,camerin2019microgels} given explicitly in the Supporting Information which contains the size $\sigma$ of the repulsive monomers as a length and the repulsion strength $\varepsilon$, as the same energy scale as for the FENE potential. Henceforth we choose $\sigma$ and $\varepsilon$ as units of length and energy. 
 
 To incorporate the thermoresponsivity of the microgel particle effectively, a further attractive bead-bead pair potential~\cite{Gnan2017} is added given by
\begin{equation}
V_\alpha\left(r\right)=
\begin{cases}
-\alpha\varepsilon & \text{if}\qquad r \leq 2^\frac{1}{6}\sigma\\
\frac{1}{2}\alpha\varepsilon\left[\cos\left(\gamma\left(\frac{r}{\sigma}\right)^2+\beta\right)-1\right] & \text{if} \qquad 2^\frac{1}{6}\sigma < r \leq R_0\sigma\\
0 & \text{otherwise}
\end{cases}
\end{equation}
with $\gamma=\pi \left(2.25-2^\frac{1}{3}\right)^{-1}$ and $\beta = 2\pi-2.25\gamma$~\cite{soddemann2001}.
Importantly, the effective attraction strength  is controlled by the parameter $\alpha$, which mimics the quality of the solvent in an implicit manner. The value $\alpha =1.55$ describes a strong attraction (relative to the bead repulsions) imitating poor solvent conditions ~\cite{Gnan2017,soddemann2001} and therefore mimicking the collapsed state of the microgel. For good solvent conditions the value $\alpha=0$ is used such that there is no attraction at all, ~\cite{Gnan2017,soddemann2001} reflecting the swollen state. In our simulations, we varied the effective attraction strength $\alpha$ between the two extreme cases in the range $0\leq \alpha \leq 1.55$ and refer to the $\alpha=0$ case as the "fully swollen state" and the $\alpha=1.55$ case as the "fully collapsed state". As the solvent quality is controlled by temperature in the experiments, $\alpha=0$ corresponds to a low temperature, below the volume phase transition of the PNiPAm microgels, and $\alpha=1.55$ to a high temperature situation above the volume phase transition. The internal architecture of the microgel particle is as in Ref.\ ~\cite{Gnan2017,camerin2019microgels,rovigatti2017internal} and  depends on the crosslinker density. In our simulations we represent each microgel particle by  a total number of $N=5500$ beads, which includes a fraction of homogeneously distributed crosslinkers. The percentage of  crosslinkers is an important parameter in our simulations and will be systematically varied between 0 and $4.5\%$. 

To mimic the effect of an air-water interface, we follow Ref.\ ~\cite{Harrer2019} and add an external potential; the potential is 
defined such that the interface normal is along the $z$ direction. Explicitly, for each bead, the external potential is a combination of an effective Lennard-Jones potential on the water side  and a steep linear potential on the air side.  We introduce a typical range $\sigma_\text{ext}$ for the effective bead-interface interaction and a shifted $z$-coordinate $\tilde{z}$ by $\tilde{z} = z -2^\frac{1}{6}\sigma_\text{ext}$; the  derivative of the potential (i.e.\ the force) is defined to be continuous at a  matching point $\tilde{z}_a<0$.
In detail, 
\begin{equation}
V_\text{ext}\left(\tilde{z}\right)=
\begin{cases}
{V}_\text{LJ}\left( \tilde{z}  \right) & \tilde{z} \geq \tilde{z}_a \\
V_\text{LJ}\left(\tilde{z}_a\right)+(\tilde{z}_a-\tilde{z})\left.\frac{dV_\text{LJ}\left(\tilde{z}\right)}{d\tilde{z}}\right|_{\tilde{z}=\tilde{z}_a} & \tilde{z} < \tilde{z}_a
\end{cases}
\label{eqn:ext_pot}
\end{equation}
with the Lennard-Jones potential
\begin{align}
{V}_\text{LJ}\left(\tilde{z}\right)=4 \varepsilon_\text{ext}\left[\left(\frac{\sigma_\text{ext}}{\tilde{z}}\right)^{12}-\left(\frac{\sigma_\text{ext}}{\tilde{z}}\right)^6\right],
\end{align}
where $\varepsilon_\text{ext}$ is an attraction energy strength. Therefore the interface tries to pin all beads  to the position $\tilde{z}=0$ where the potential is minimal. Physically this attraction towards the interface results from surface tension reduction by reducing the bare air-water interfacial area when a bead is adsorbed at the interface. The large difference in the bead chemical potential between the air and the water phase is reflected by the steep increase of $V_\text{ext}\left(\tilde{z}\right)$ for $\tilde{z}<\tilde{z}_a$. In the following we have chosen $\varepsilon_\text{ext} = 5.5\varepsilon$, to ensure a sufficiently strong adsorption strength towards the interface, and $\sigma_\text{ext}=0.5 \sigma$ corresponding to a relatively deep and stiff minimum around $\tilde{z}=0$. A harmonic expansion around the origin $\tilde{z}=0$ yields a large  spring constant of $22\epsilon/\sigma^2$ which is enforcing a bead monolayer as observed by the very thin corona formed by polymer chains expanding at the liquid interface. Finally we have chosen the matching position $z_a$, i.e.\ the point where the Lennard-Jones behavior changes to  a constant force, as  $\tilde{z}_a=-0.01\sigma$. 

We simulate the bead dynamics as Brownian dynamics with an implicit solvent, whereby the short-time bead self-diffusion coefficient $D_0$ sets a characteristic Brownian time scale $\tau_B = D_0/\sigma^2$. The latter defines our unit of time in the following. A finite time step of $\Delta t=0.00005\tau_B$ is used to integrate the equations of motion with an Euler forward scheme.
All of the Brownian dynamics simulations are performed with the HOOMD-Blue package ~\cite{anderson2020hoomd} and are visualized by OVITO~\cite{stukowski2009visualization}. 

In our modelling it is important to distinguish between a solvent bath temperature $T^*$ -- which sets the Brownian fluctuations -- and the implicit temperature influence on the effective bead-bead attraction strength $\alpha$. Since the temperature change is small compared to the absolute room temperature, we have fixed the solvent bath temperature to $k_\text{B}T^*=\varepsilon$ throughout all of our simulations but changed solely  the effective attraction strength $\alpha$, in agreement with typical protocols in the literature ~\cite{Gnan2017,rovigatti2017internal}.

Our simulation protocol is as follows: first we equilibrate the microgel particle in the bulk (i.e.\ in the absence of the interface external potential (\ref{eqn:ext_pot})) and thus gain an equilibrated initial bead configuration. This is done separately for the "fully swollen state" with $\alpha=0$ and for the "fully collapsed state" with $\alpha=1.55$. We then instantaneously expose the initial bead configuration to the potential (\ref{eqn:ext_pot}) such that the $z$-coordinate of the center of mass of the particle is a distance of $10\sigma$ apart from the interface position at $\tilde{z}=0$. We then relax the system for  a long waiting time $t_w$ of typically $5\cdot 10^3\tau_B$. The relaxed configuration is -- in the presence of the interface -- subject to a sudden change in the attraction parameter $\alpha$ from 0 to 1.55 or, respectively from 1.55 to 0. Physically this means that the solvent quality temperature is abruptly changed from high to low (or vice versa). We then allow the system to undergo relaxation for another waiting time; this relaxation process is referred to as "collapsing" (if $\alpha$ has been suddenly increased), or  as "swelling" (if $\alpha$ has been suddenly  decreased). Finally we reverse $\alpha$ to its initial value, thus establishing one "cycle".
After a third waiting time the resulting configuration is compared to the first configuration at the same $\alpha$ relaxed at the interface from its initial bulk state. If there is a significant difference in the extent of the configuration, the system is called hysteretic. The protocol is repeated several  times leading to a "cycling process" which is composed of alternating "swelling" and "collapsing" processes. Appropriate averages are taken to ensure that the behavior does not suffer from peculiarities of the initial configuration. 

The diagnostics to identify hysteresis is done via monitoring the lateral radius of gyration, $R_\text{lat}(t)$, as a function of time $t$. The lateral (or projected) radius of gyration is defined as
\begin{equation}
R^2_\text{lat}(t)=\frac{1}{N}\sum_{i=1}^N(x_i(t)-X_0(t))^2+(y_i(t)-Y_0(t))^2
\end{equation}
with $\vec{r}_i(t)=(x(t), y(t),z(t))$ as the location of the bead $i$, and ${\vec R}_0(t) = (X_0(t), Y_0(t),Z_0(t))= \frac{1}{N} \sum_{i=1}^N \vec{r}_i(t)$ is the instantaneous center of the microgel particle. Finally, we can also infer the lateral osmotic pressure from the fraction of beads in the interfacial region. Here, by definition, a bead is in the interfacial region  if its $\tilde{z}$-coordinate lies between $-0.5\sigma$ and $0.5\sigma$. We also analyze the internal core-corona structure in more detail by using a "hull parameter" $\Delta (r)$, which brings us also into a position to extract a core radius $R_c$ and an outer corona radius. Details of the procedure are described in the Supporting Information.
 
\subsection{Experiments}
PNiPAm microgels crosslinked by N,N'-Methylenebis(acrylamide) with different crosslinking densities are synthesized by precipitation polymerisation according to a previously published protocol~\cite{rey2017interfacial}. 
The cycling experiments involving the differently crosslinked microgels are conducted in a similar fashion as developed in an earlier publication~\cite{Harrer2019}. When starting from the collapsed state, 0.05 w\% microgels dispersed in an ethanol/water mixture (1:1) with a temperature of $70~^\circ\text{C}$ are spread at the air/water interface on a Langmuir-Blodgett trough (KSV Nima), which is preheated to the initial temperature of $60~^\circ\text{C}$. Similarly, when starting from the swollen state, the microgel dispersion is spread at room temperature. The trough is heated using a thermostat by increasing the temperature by $10~^\circ\text{C}$ every 10 minutes. Conversely, the trough is passively cooled by switching off the thermostat. The interfacial arrangement as a function of temperature is then deposited onto a silicon wafer (0.5 x 10 $\mathrm{cm}^2$) lifted at 0.1 mm/min through the interface, while the corresponding surface pressure is measured using the Wilhelmy plate method. The microgels are characterized ex-situ by image analysis of scanning electron microscopy (Zeiss Gemini 500) images of the transferred interfacial layer. From these images, the size of the microgel core is quantified by a custom-written Matlab software~\cite{Harrer2019}.

\section{\label{III}Results and Discussion} 
\subsection{Non-hysteretic microgel swelling in the bulk}
\begin{figure}[h]
\centering
\includegraphics[height=4.25cm]{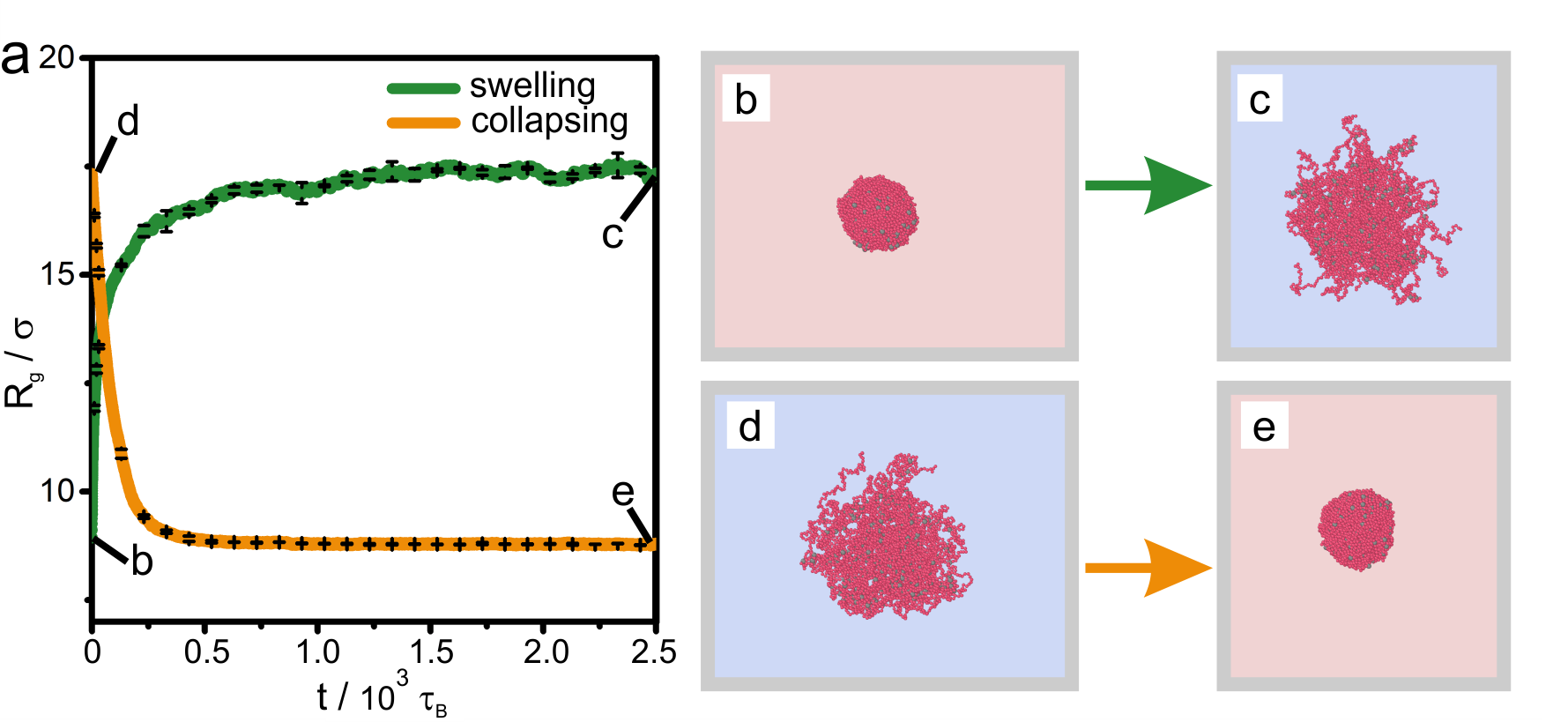}
\caption{Volume phase transition of a microgel in bulk. Radius of gyration $R_g$ (in terms of the bead size $\sigma$) as a function of time $t$ for a microgel particle in the bulk for both a collapsing process (yellow line)  and a swelling process (green line) with the corresponding error bars in black as obtained by simulation. The processes start from equilibrated configurations (shown as typical simulation snapshots b) and d)). Typical snapshots c) and e)  after a long simulation time of $2500\tau_B$ are also given. There is no notable hysteresis in the bulk. The data are obtained  for a crosslinking density of $4.5\%$.}
\label{fig:bulk}
\end{figure}

First, as a reference, we simulate the bulk behavior of a microgel. Figure \ref{fig:bulk} shows the time evolution of the radius of gyration during swelling and collapsing together with typical initial and final simulation snapshots before and after the swelling and collapsing processes. The initial configurations (marked with b) and d)) are equilibrated for a long relaxation time of typically $5000 \tau_\text{B}$, and are therefore practically fully relaxed. After a sudden increase (or decrease) of effective attraction $\alpha$ (mimicking  an increase or decrease the temperature, respectively) the radius saturates quickly (i.e. within a relaxation time of roughly $1000  \tau_B$) indicating equilibration. After a long relaxation for a simulation time of about $2500 \tau_B$ two final configurations marked with c) and e) in Figure \ref{fig:bulk}  are reached. These two final configurations c) and e) are similar to the initial conditions d) and b) in Figure \ref{fig:bulk}, i.e.\ both swollen and both collapsed configurations exhibit a similar radius of gyration. Importantly, this implies that swelling and subsequent collapsing are completely reversible in this bulk experiment on the accessible  time scale of the simulation, or in other terms, there is no hysteresis. This non-hysteretic bulk  behavior in size is in agreement with experiments ~\cite{Kratz2001}.  Once the reversible behavior of the volume phase transition is established, one can trivially add a cycle of  subsequent swelling and collapsing processes, which always results in an equilibrated state. Likewise, one can perform a multi-cycle process by adding many subsequent swelling and collapsing processes. Again one finds reversible behavior such that after each cycling process the system is in the same equilibrated state. The latter feature will change significantly at interfaces which we shall show next.

\begin{figure*}[h]
\centering
\includegraphics[height=4cm]{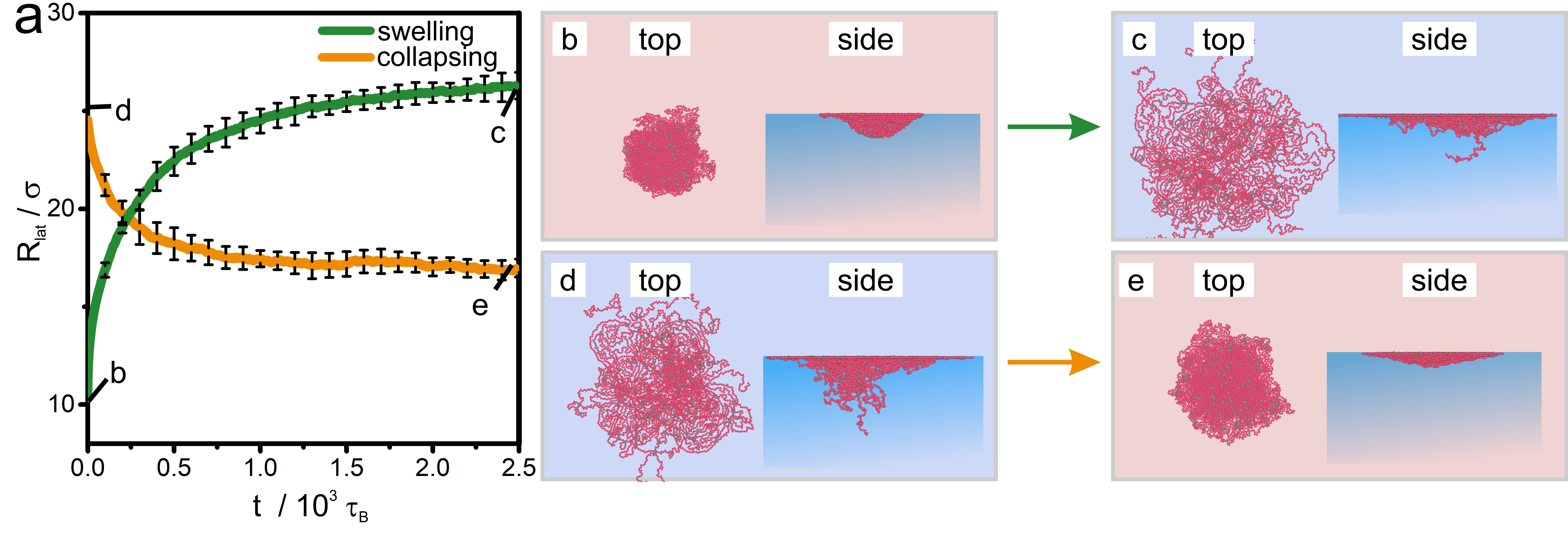}
\caption{Volume phase transition of a microgel at an interface. Lateral radius of gyration $R_\text{lat}$ (in terms of the bead size $\sigma$) is shown as a function of time $t$ for a microgel at the liquid interface for both a collapsing process (yellow line) and a swelling process (green line) with the corresponding error bars in black as obtained by simulation. The processes are started from  configurations relaxed at the interface for a total time of $5000\tau_B$ (shown as typical simulation snapshots b) and d)). Typical snapshots c) and e)  after a long relaxation time of $2500\tau_B$ are also given. There is a notable hysteresis for the collapsing but not for the swelling process. The data are obtained for a crosslinking density of 4.5\%}
\label{fig:interface}
\end{figure*}

\subsection{Liquid interface-induced hysteresis of microgel morphology}
\subsubsection{The swelling and collapsing process at the interface}
Next, a microgel with an equilibrated bulk configuration is placed at the interface in silico (as described in Materials and Methods) and relaxed there for a long initial relaxation time  of $5000 \tau_B$. Subsequently, a swelling and collapsing process is initiated by an instantaneous change in the solvent quality, encoded in the effective attraction $\alpha$. The  evolution of the lateral radius of gyration following this change of $\alpha$ is presented in Figure \ref{fig:interface} for both a swelling and a collapsing event. Similar to Figure \ref{fig:bulk}, typical initial and final snapshots are given in Figure \ref{fig:interface} both with a top view onto the microgel and a lateral view from the side. The side view clearly shows the oblate core-corona morphology ~\cite{geisel2012unraveling,camerin2019microgels} of the microgel, which is induced by lateral stretching caused by the reduction of the bare water-air interfacial tension. This is also accompanied by the fact that the lateral radius of gyration is significantly larger than its bulked value shown in Figure \ref{fig:bulk}. The important message taken from Figure \ref{fig:interface} is the presence of a hysteresis, observed from the difference in the collapsed dimensions for the two phase transition scenarios. In particular, the collapsed state e), reached after a relaxation time of $2500\tau$ from an initially swollen microgel at the interface, is much less contracted than the starting collapsed configuration b) from the initial interfacial adsorption of a bulk collapsed microgel. This difference is clearly evidenced by the different lateral radii of gyration, and is also consistent with previous experimental findings ~\cite{Harrer2019,Schulte2019,bochenek2020temperature}. 

The  hysteretic behavior found for the volume phase transition at the interface in the collapsing process is in  contrast to the swelling process where the swollen end state c) practically exhibits the same lateral radius of gyration as the initial state d) of a microgel adsorbed to the interface from a swollen bulk conformation. The clue to understand this difference lies in the nature of the starting state b). In the collapsed state at the interface,the equilibration is kinetically hindered by the strong attractive interactions between the beads. Therefore we hypothesize that they are kinetically trapped  since they cannot escape quickly to relax all  constraints. If a swelling process is induced by improving the solvent quality, these constraints are released and the microgel relaxes faster to its equilibrium state. Again we emphasize that the allotted relaxation time at the interface for the collapsed state is long as compared to the typical time for swelling and collapsing in the simulation but not enough to achieve full equilibration.  This "explosion" in relaxation time scales for interfacially collapsed states is in accordance with the experimental findings in literature ~\cite{Harrer2019,Schulte2019,bochenek2020temperature}. 

\subsubsection{Temperature cycling}
\begin{figure}[h]
\centering 
\includegraphics[height=4cm]{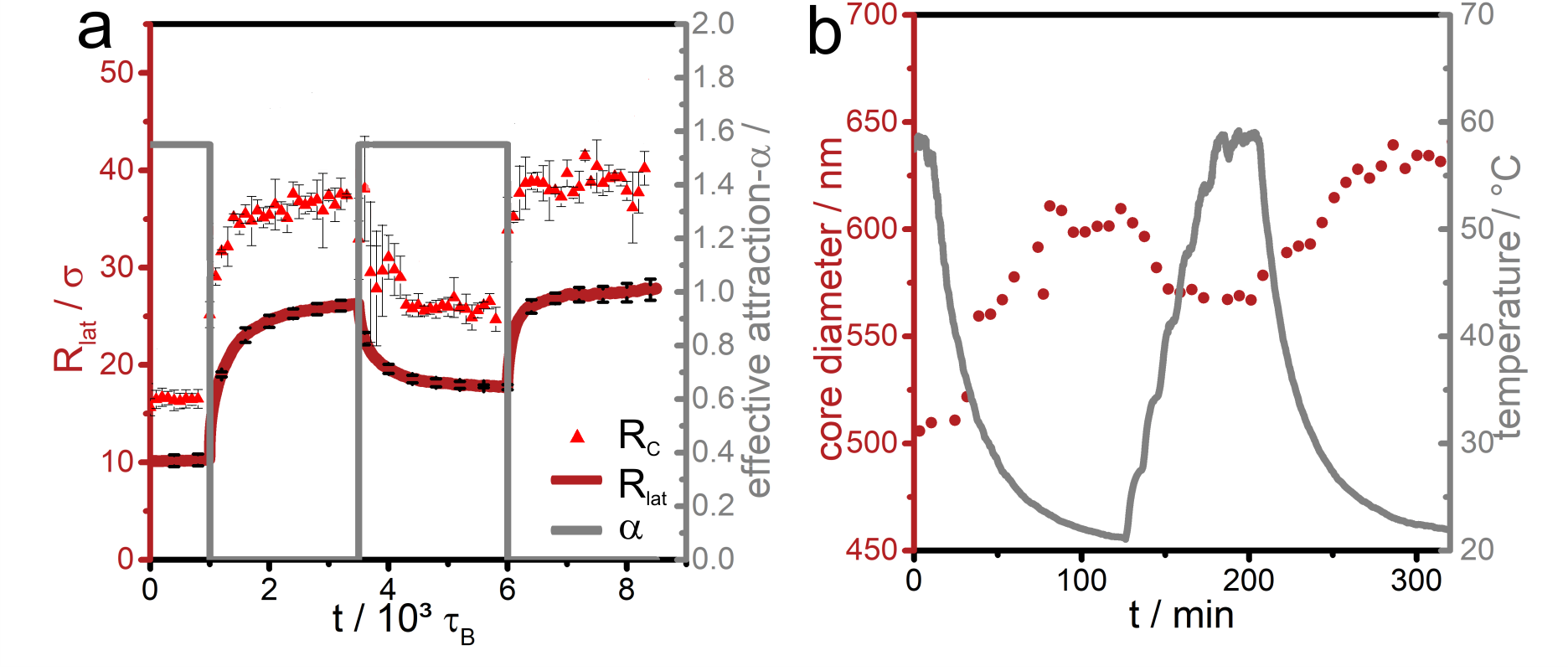}
\caption{Temperature cycling of a microgel at an interface. Cycling process of swelling and collapsing for a microgel particle initially collapsed
 at the interface. a) simulation data and b) experimental data.
A hysteresis within the first temperature cycle can be clearly observed for both simulation and experiment.
The simulation data are obtained  for a crosslinking density of $4.5\%$ and the experimental data for a crosslinking density of $5\%$. The corresponding effective attraction parameter $\alpha$ and the solvent temperature are also shown as a function of time in a) and b) to illustrate the cyclic nature of the process. Two different length scales are shown in the simulation data in a) which give the same qualitative trend,
namely the core size $R_c$ and the lateral radius of gyration $R_\text{lat}$.}
\label{fig:temp_cycling}
\end{figure}

We now go one step further and study a periodic sequence of swelling and subsequent collapsing processes in a cycling way. In particular, this procedure allows investigation whether the swollen state arising by swelling from an initial collapsed state is hysteretic upon further periodic decrease and increase of solvent quality (or temperature).
Simulation data for such a cycling process are given in Figure \ref{fig:temp_cycling}a. These data show an almost hysteresis-free behavior after one cycle.
This gives evidence that the hysteretic behavior is mainly attributed to the collapsed initial state b). Once this adsorbed, collapsed microgel is swollen, the resulting interfacial state can almost reversibly be collapsed and swollen again without showing any further hysteresis in radius of gyration.

The hysteretic behavior is reflected in the two length scales characterizing the microgel particle,
namely the lateral radius of gyration $R_\text{lat}$ and the core radius $R_c$ as documented in Figure \ref{fig:temp_cycling}a.
Indeed our analysis of the morphology by the hull parameter $\Delta(r)$ (Fig.\ 7, 
Supporting Information) reveals a clear distinction between core and corona for the different states in the cycling process and documents that our simulation scheme reproduces the fried-egg structure of a microgel at the interface~\cite{Geisel2012,Style2015,Mehrabian2016,Zielinska2016,Rey2020}. The core region is much more compressed due to the interfacial attraction to the beads. Therefore the effect of kinetic arrest in the core is much more amplified when the particle is brought to the interface
as compared to the bulk situation. This gives a clue to understand the underlying reason for the hysteretic behavior at the interface.

We compare the hysteresis observed in simulation with experimental data,  obtained for an interfacial assembly of microgels on the air/water interface of a Langmuir trough exposed to changes in the temperature of the water subphase. the diameter of the microgel cores are quantified by image analysis of scanning electron microscopy images after transfer to a solid substrate. Interestingly, the experimental data shows a similar hysteretic behavior, evidenced by a change in core diameter between the initial adsorption in collapsed state and the first temperature cycling as seen in  Figure \ref{fig:temp_cycling}b. However, in the experiments the change of solvent
quality (temperature) is not stepwise but smoothened due to experimental constraints (compare the gray line in Figure \ref{fig:temp_cycling}b to the sharp jump of the
effective attraction $\alpha$ shown along with the simulation data in Figure \ref{fig:temp_cycling}a).

\subsection{Impact of the degree of crosslinking}

\subsubsection{Impact of the degree of crosslinking on the hysteresis}
\begin{figure}[h]
\centering
\includegraphics[height=7.5cm]{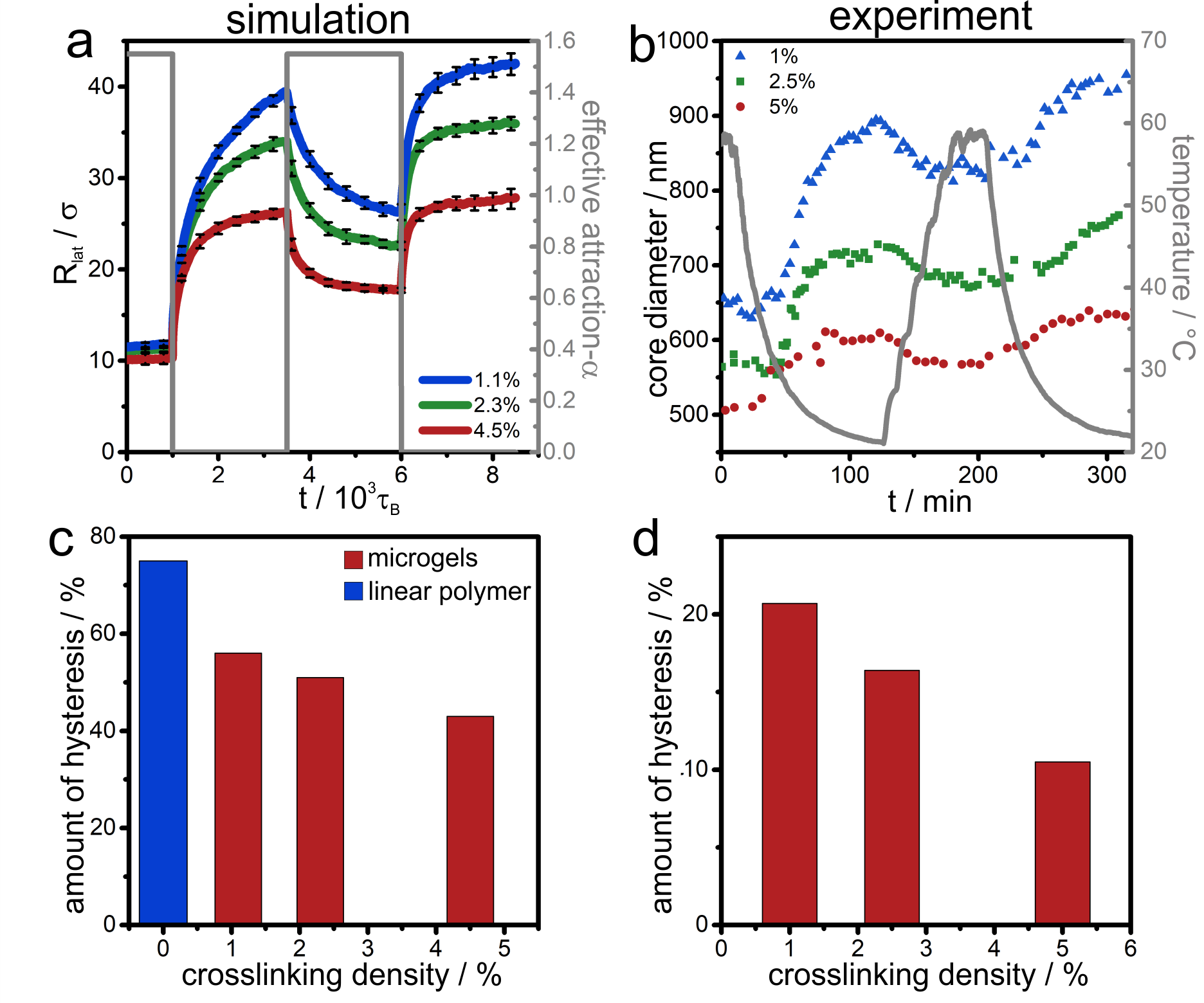}
\caption{Temperature cycling for different crosslinking densities. Cycling process of swelling and collapsing for microgel particles with different crosslinking densities initially collapsed at the interface: a) simulation data and b) experimental data. Hysteresis within one cycle can be clearly observed for both simulation and experiment. On the simulation side the lateral radius of gyration $R_\text{lat}$ is used for comparison. The corresponding effective attraction parameter $\alpha$ and the solvent temperature are also shown as a function of time in a) and b) to illustrate the cyclic nature of the process. The relative amount of hysteresis as a function of the crosslinking density: c) simulation, d) experimental data. Simulation data for the special case of a linear polymer  are added as well for comparison (blue). }
\label{fig:variedcrosslinking}
\end{figure}

We now investigate the hysteretic behavior as a function of the degree of crosslinking. In Figure \ref{fig:variedcrosslinking}
the interfacial volume phase transition upon temperature cycling  is compared  for three different crosslinking densities in  simulation and experiment. The relative amount of hysteresis is quantified as $1-\cal {R}$, where $\cal R$ is the
ratio between the two microgel radii before and after the first
cycling process. For the simulations, the ratio $\cal R$ is  taken as  $R_{lat}(t= 1000\tau_B)/R_{lat}(t= 6000\tau_B)$,
see Figure \ref{fig:variedcrosslinking}a, and for the experiments
we take for $\cal R$ the ratio of the two radii at times $210\,\text{min}$ and $0\,\text{min}$, see Figure \ref{fig:variedcrosslinking} b. In the absence of hysteresis, the two radii coincide such that
$\cal R$ is one and the amount of hysteresis vanishes. 

Both experimental and simulation data show that the relative amount of hysteresis decreases with increasing crosslinking density (\ref{fig:variedcrosslinking}c,d).  We rationalize this behavior based on an intuitive argument stemming from the kinetics of the initially trapped state. For a more connected polymer network the microgel is more resistant against the stretching effect of the interface and therefore it is stretched out less.
This explains why the hysteresis is smaller for higher crosslinking density. Conversely, if the degree of crosslinking is small,
 long dangling chains can be much more entangled and therefore contribute much more to the amount of hysteresis.
Indeed in the extreme case of a single linear chain, additional simulation and experimental data indicate
that the relative amount of hysteresis is even higher than the values found for the lowest crosslinking density of one percent; for simulation data see again Figure \ref{fig:variedcrosslinking}c. The number of beads in the simulation of the linear
chain is identical to that contained in the microgel particle, namely 5500 beads.

\begin{figure*}[h]
\centering
\includegraphics[height=8cm]{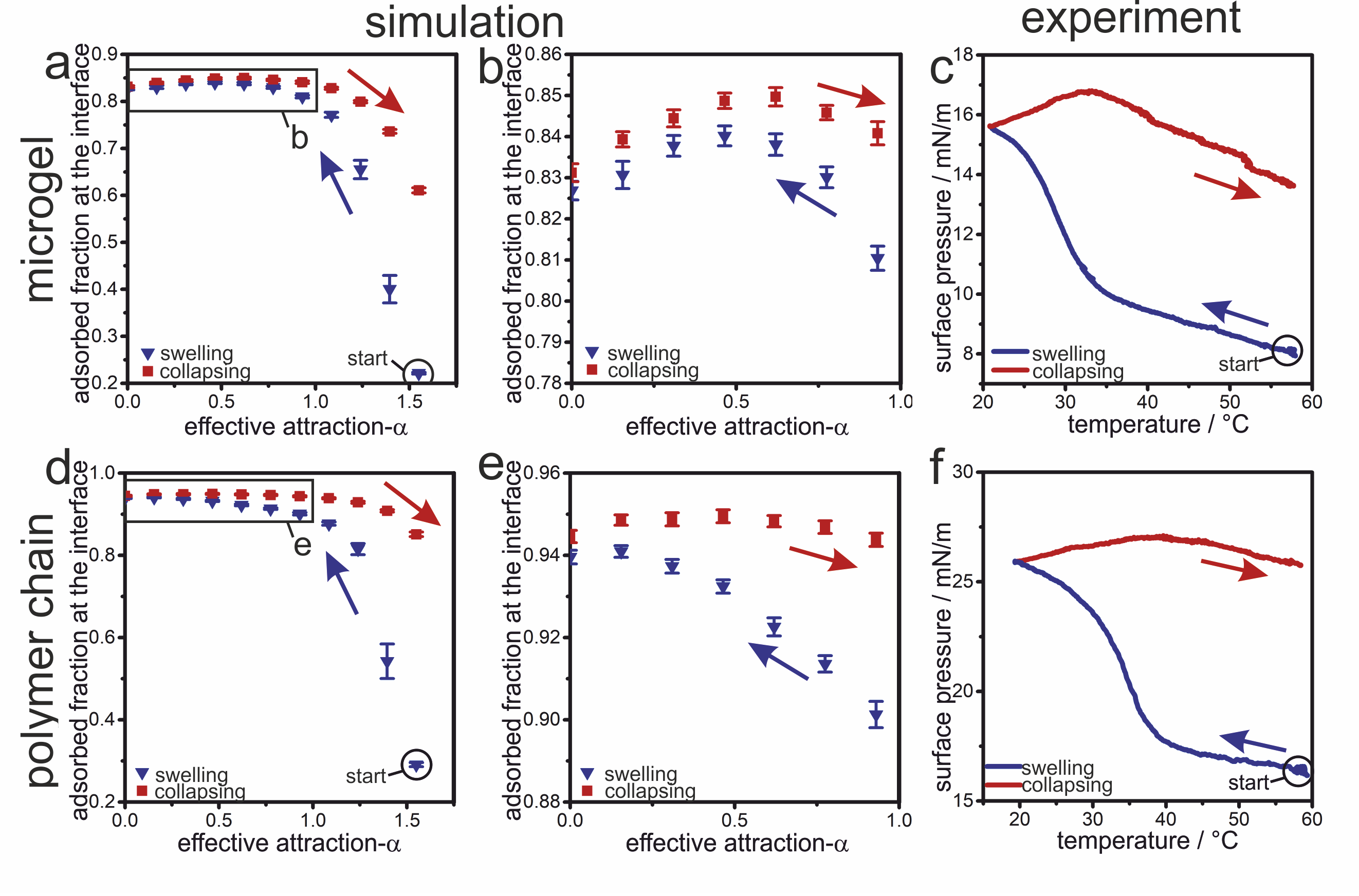}
\caption{Surface pressure cycle for different crosslinking densities. Evolution of surface pressure during the volume phase transition at interfaces, for a microgel and a linear polymer chain. a) simulation data for a microgel ($2.3\%$ crosslinking density). b) a zoom-in for the microgel simulation data for effective attractions $\alpha$ between 0 and 1. c) experimental data for a PNiPAm microgel ($2.5\%$ crosslinking density). d) simulation data for a linear polymer chain. e) a zoom-in for the linear polymer chain for effective attractions $\alpha$ between 0 and 1. f) experimental data for a linear PNiPAm  polymer chain.  A hysteretic behaviour of the volume phase transition is observed for both microgel and linear polymer chain in simulation and experiment.  }
\label{fig:surfacepressure}
\end{figure*}

\subsubsection{Comparison of a microgel and a linear chain in terms of the surface pressure}
We now consider the surface pressure of an interfacial microgel layer as a function of temperature, which can be measured in experiments. Care has be taken in the interpretation of the effect of microgels on the surface pressure, since the air/water surface tension also changes with temperature. As the latter shows a monotonic decrease, we attribute any deviation from such a linear behavior to the microgel layer and therefore subtract the change in surface pressure of water from the plotted data. Therefore, we refrain here from a full quantitative comparison between experiment and simulation
but only consider qualitative trends.  

In the experimental system, the decrease in surface tension in the presence of microgels (or, the increase in surface pressure) relates to the surface density of PNiPAm chains adsorbed to the air/water interface. In simulations, the corresponding quantity is the surface density of adsorbed beads, which is directly accessible.
Figure \ref{fig:surfacepressure}a and c compares the evolution of the adsorbed microgel fraction determined in simulation with the surface pressure measured in experiment as a function of the effective attraction $\alpha$ or the temperature, respectively.  The data are obtained by starting with an interfacially adsorbed microgel with high effective attractions $\alpha$ resp.\ high temperature $T$ (solid points). This initially collapsed microgel is then slowly swollen at the interface by a decrease the attraction (resp.\ temperature),  as shown by blue points/blue line.
The swelling increases the area occupied by the microgel (see above), and thus decreases the surface tension - or increases the surface pressure in experiment (\ref{fig:surfacepressure}c).
Concomitantly, the swelling causes more beads to adsorb to the interface as the system passes through the volume phase transition (\ref{fig:surfacepressure}a,b).
 
Notably, the maximal number of adsorbed beads occurs closely at
$\alpha \approx 0.6$ where the bulk collapse transition happens~\cite{Gnan2017, rovigatti2017internal} and is thus correlated to the bulk volume transition. Interestingly, in the simulation data, a slight {\it decrease\/} of the adsorbed beads is absorbed when the attraction vanishes. This is due to the fact that parts of the chains go back into the bulk due to entropic reasons.
The swollen microgels are subsequently collapsed (or heated) again. The red data points/red line in Figure \ref{fig:surfacepressure}a-c show the associated evolution of the adsorbed fraction of microgel beads (simulations) and surface pressure (experiments), respectively. In both experiment and simulation, a hysteresis is observed, in agreement with the data shown before. The fraction of  adsorbed beads and the surface pressure increases for this increase in $\alpha$ or temperature, compared to the initial values at low temperature culminating in a maximum at a temperature which roughly coincides with the bulk volume transition. We interpret this maximum as a joint effect of the interfacial attraction to the beads and the effective attraction between the beads. When the bulk volume transition temperature is reached from below, the bulk attraction wins and drags beads from the interface into the bulk.
We remark that upon cooling a similar maximum  is found in the simulations but not in the experiments; the reason for this slight discrepancy remains unclear.

Finally, as a reference,  we also show data for a linear polymer chain in Figure \ref{fig:surfacepressure}c-f.
The linear chain shows similar trends in the surface pressure and the fraction of the adsorbed monomers as the microgel, but
there is no  maximum in the adsorbed bead fraction during the swelling process. Thus the curves coincide nicely.

\section{\label{V}Conclusions}

We used monomer-resolved computer simulations and interfacial experiments to investigate the volume phase transition of PNiPAm microgels in bulk and adsorbed to an interface. Our results underline that the presence of an interface significantly changes the volume phase transition. Particularly, we
 found a significant hysteretic behavior for microgels undergoing the phase transition at the interface. A microgel adsorbed in the collapsed state to the interface does not return to its initial configuration when subjected to a temperature cycle. Instead, it relaxes into a more stretched configuration. An initially swollen microgel, however, undergoes reversible transitions between collapsed and swollen states upon temperature cycling. We therefore attribute the hysteresis effect of the collapsed microgel at the interface to a kinetically trapped initial state
which can be released by swelling the microgel. We find that the hysteresis is more pronounced for weaker degree
of crosslinking and is even observed for linear PNiPAm chains. 

Our results demonstrate that it is possible to model complex polymer-interface phenomena 
via a comparably simple model that balances internal and interfacial attractions. This modelling approach may therefore also be transferred to other stimuli-responsive polymer systems, curved interfaces, or more crowded systems formed by multiple, overlapping interfacial microgels.


\section*{Conflicts of interest}
There are no conflicts to declare.

\section*{Acknowledgements}
N.V., H.L. and L.M.C.J. acknowledge funding from the Deutsche Forschungsgemeinschaft (DFG) under grant numbers VO 1824/8-1 and LO 418/22-1, respectively. N.V. also acknowledges support by the Interdisciplinary Center for Functional Particle Systems (FPS). J.K. thanks Jens Grauer for helpful discussions.


\section{Supporting Information}
\subsection{Bead and bond potentials}
The bead-bead interaction is modeled by a Weeks-Chandler-Andersen~\cite{weeks1971} potential
\begin{equation}
V_\text{WCA}\left(r\right)=
\begin{cases}
4\varepsilon \left[\left(\frac{\sigma}{r}\right)^{12}-\left(\frac{\sigma}{r}\right)^6\right]+\varepsilon &  \text{if} \qquad r \leq 2^\frac{1}{6}\sigma\\
0 & \text{otherwise}
\end{cases}
\end{equation}
where $r$ is the radial distance between two beads,  $\sigma$ represents the bead size  and $\epsilon$ is the  strength of repulsion. The bead-connecting covalent bonds are described by a finite-extensible-nonlinear-elastic (FENE) ~\cite{Gnan2017,camerin2019microgels} potential
\begin{equation}
V_\text{FENE}\left(r\right)=
\begin{cases}
- \tilde{k}_\text{F}\tilde{R}_0^2\ln\left(1-\left(\frac{r}{\tilde{R}_0}\right)^2\right)&\text{if} \qquad r<\tilde{R}_0 \\
0 & \text{otherwise}
\end{cases}
\end{equation}
with $\tilde{k}_\text{F}=15\varepsilon/\sigma^2$ an effective spring constant and $\tilde{R}_0=1.5\sigma$ the maximal bond expansion.

\subsection{Core-corona structure}
We characterize the internal core-corona structure of the microgel by a geometric analysis of the bead configurations. The main idea is as follows: we consider all bead positions projected to the $xy$-plane which have a distance less than a prescribed $r$ from the (stationary) microgel center ${\vec R}_0$, i.e.\ all bead positions which fulfill $(x_i(t)-X_0(t))^2+(y_i(t)-Y_0(t))^2<r$. In the corona region we expect that the set of bead positions exhibits a relatively rough boundary while in the core region the boundary is fairly smooth.

To quantify this further we consider the convex hull around the set of bead positions which provides a contour length $L_{convex}(r)$. We also define a {\it concave\/} hull around the same bead positions and compare its contour length $L_{concave}(r)$ to $L_{convex}(r)$. Clearly, $L_{concave}(r) \geq L_{convex}(r)$. In contrast to the convex contour, the precise definition of the concave hull is not unique. In detail, for the calculation of the concave hull we employed the algorithm in Ref.\ ~\cite{Moreira2007} with the so-called $k$-nearest-neighbour approach where we used $k=20$.
 
The relative difference in the two contour lengths defines a parameter $\Delta (r)$
\begin{equation}
\Delta(r)=\frac{L_\text{concave}(r)-L_\text{convex}(r)}{L_\text{concave}(r)}
\end{equation}
The spatial dependence of  $\Delta (r)$ contains valuable information about the core size $R_\text{c}$ and the corona size $R_\text{co}$. Inside the core $\Delta$ is small while it increases for increasing $r$ until it saturates with a value $\Delta_\infty$ of the order of one, concomitant with a "rugged" concave hull. We can therefore extract the core and corona size approximatively by studying the behavior of $\Delta (r)$ near its inflection point at $r=r^*$ defined by the condition $d^2 \Delta (r)/dr^2|_{r=r^*}=0$. Considering the tangent ${\tilde t}(r)$ through the inflection point given by the linear relation ${\tilde t}(r)=\Delta (r^*) + d \Delta (r)/dr|_{r=r^*} \times (r-r^*)$, we define the core size $R_\text{c}$ by the intersection of the tangent with the $r$-axis, i.e.\ by the condition ${\tilde t}(R_\text{c})=0$. The corona size is defined by the distance where the tangent reaches the saturation value $\Delta_\infty$, i.e.\ by the relation ${\tilde t}(R_\text{co})=\Delta_\infty$.

\begin{figure}[h]
\includegraphics[height=11.5cm]{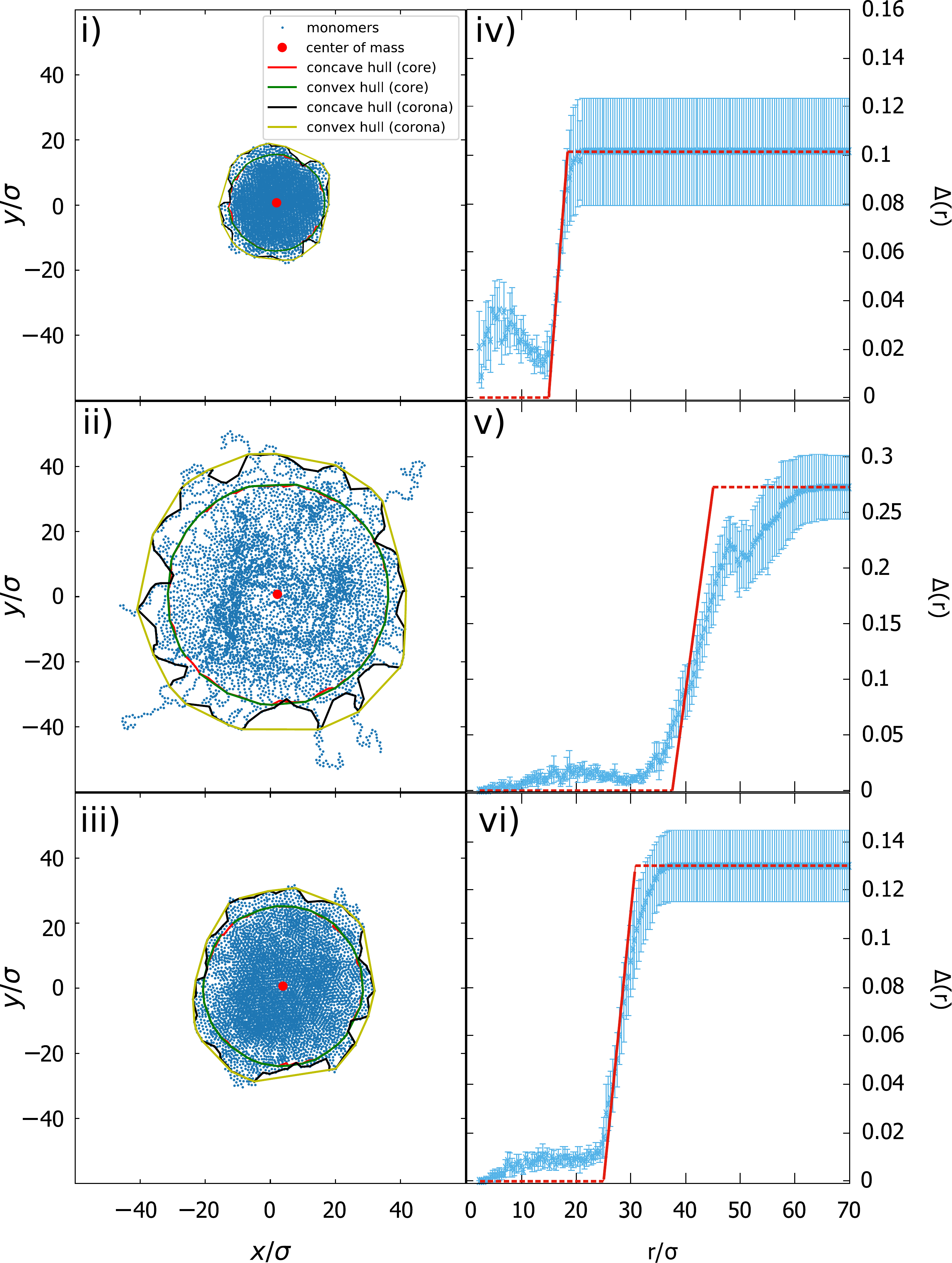}
\caption{i)-iii) Typical bead configurations for a crosslinking density of $4.5\%$ at the interface (top view) for the initial collapsed state i), the swollen state ii) and the collapsed state after one cycle iii). Different hulls for the core and  corona are indicated in different colors (see legend). The corresponding $\Delta(r)$-profiles are shown in iv), v) and vi) with error bars. The tangent on the $\Delta(r)$-profile is shown as a full red line, the intersection point with the dashed lines indicates the core and corona size.}
\label{sup:rel_hull_dif1}
\end{figure}

Concrete data for a typical microgel snapshot along with the profiles for $\Delta (r)$ are presented in figure \ref{sup:rel_hull_dif1}i)-iii) corresponding to three different situations: i) initially collapsed state of a microgel particle brought to the interface from the bulk, ii) swollen microgel particle at the interface after the first temperature change, iii) collapsed state after one full cycle. The associated concave and convex hulls for the  inner and outer part are indicated in different colors.  Moreover, in figure \ref{sup:rel_hull_dif1}iv)-vi) the corresponding profiles $\Delta (r)$ are presented together with  the tangent ${\tilde t}(r)$ (full red line) and the determination of the core size $R_\text{c}$ and the corona size $R_\text{co}$.

\balance


\bibliography{manuscript.bib}
\bibliographystyle{rsc} 
\end{document}